\documentclass{vgtc}                          % final (conference style)
\ifpdf%                                % if we use pdflatex
  \pdfoutput=1\relax                   % create PDFs from pdfLaTeX
  \usepackage{graphicx}                % allow us to embed graphics files
  \DeclareGraphicsExtensions{.pdf,.png,.jpg,.jpeg} % for pdflatex we expect .pdf, .png, or .jpg files
\else%                                 % else we use pure latex
  \usepackage{graphicx}                % allow us to embed graphics files
  \DeclareGraphicsExtensions{.eps}     % for pure latex we expect eps files
\fi%

\graphicspath{{figures/}{pictures/}{images/}{./}} % where to search for the images

\usepackage{microtype}                 % use micro-typography (slightly more compact, better to read)
\PassOptionsToPackage{warn}{textcomp}  % to address font issues with \textrightarrow
\usepackage{textcomp}                  % use better special symbols
\usepackage{mathptmx}                  % use matching math font
\usepackage{times}                     % we use Times as the main font
\usepackage{cite}                      % needed to automatically sort the references
\usepackage{tabu}                      % only used for the table example
\usepackage{booktabs}                  % only used for the table example
\usepackage{xurl}
\usepackage{paralist}

\newcommand{\eg}{e.g.}

\newcommand{\rv}[1]{\textcolor{black}{#1}}

\onlineid{1133}

\vgtccategory{Theory or Model}

\vgtcinsertpkg

\title{Toward Systematic Considerations of Missingness in Visual Analytics}

\newcommand*\samethanks[1][\value{footnote}]{\footnotemark[#1]}

\author{
\parbox{0.96\textwidth}{Maoyuan Sun\thanks{e-mail: \{smaoyuan, myue, lyujun\}@niu.edu.} \hspace{5mm} Yue Ma\samethanks  \hspace{5mm} Yuanxin Wang\thanks{e-mail: \{y2587wang, jianzhao\}@uwaterloo.ca.} \hspace{5mm} Tianyi Li\thanks{e-mail: li4251@purdue.edu} \hspace{5mm} Jian Zhao$^\dagger$ \hspace{5mm} Yujun Liu$^{*}$ \hspace{5mm} Ping-Shou Zhong\thanks{e-mail: pszhong@uic.edu}}\\
\parbox{0.92\textwidth}{\scriptsize{$^{*}$Northern Illinois University \hspace{0.07\textwidth} $^\dagger$University of Waterloo} \hspace{0.07\textwidth} $^\ddagger$Purdue University \hspace{0.07\textwidth} $^\S$University of Illinois Chicago} \\
}

\abstract{%
Data-driven decision making has been a common task in today’s big data era, from simple choices such as finding a fast way to drive home, to complex decisions on medical treatment.
It is often supported by visual analytics.
For various reasons (e.g., system failure, interrupted network, intentional information hiding, or bias), visual analytics for sensemaking of data involves missingness (e.g., data loss and incomplete analysis), which impacts human decisions.
For example, missing data can cost a business millions of dollars, and failing to recognize key evidence can put an innocent person in jail.
Being aware of missingness is critical to avoid such catastrophes.
To fulfill this, as an initial step, we consider missingness in visual analytics from two aspects: \textit{data-centric} and \textit{human-centric}.
The former emphasizes missingness in three data-related categories: \textit{data composition}, \textit{data relationship}, and \textit{data usage}. 
The latter focuses on the human-perceived missingness at three levels: \textit{observed}-level, \textit{inferred}-level, and \textit{ignored}-level.
Based on them, we discuss possible roles of visualizations for handling missingness, and conclude our discussion with future research opportunities.
} % end of abstract


\begin{document}

\firstsection{Introduction}

\maketitle

For various reasons (\eg, system failures, network problems, intentional information hiding, or bias), a human sensemaking process involves \textit{missingness}, such as missing data, biased data selection \cite{wall2017warning}, or partially finished analyses.
It impacts human decisions and may cause severe consequences.
For example, 
%using a heart rate chart with missing data may result in a misdiagnosis of cardiovascular disease, which could cost a patient's life.
missing data costs millions of dollars per year in business \cite{forbes17};
%%%%%%%%%%%%%%% this was addressed by adding a business example %%%%%%%%%%%%%%%
%\jian{See my comments in the abstract}
and due to failing to notice incomplete evidence, John Bunn was falsely convicted of murder \cite{AlexandraKing}, and Sunil Tripathi was wrongfully accused as a suspect in the Boston Marathon bombing on social media \cite{IreneOgrodnik}.
%%%%%%%%%%%%%%% this was addressed by revising the boston one %%%%%%%%%%%%%%%
%\jian{It is not clear these cases are due to missing data, especially for the Marathon one.}
These tragedies lead us to question: what can be missing in analytics; can we design techniques to prevent people from falling into traps of such missingness?

While investigating missingness in analytics remains an elusive skill and an understudied task, a fair amount of effort has been put in a focused direction: missing data estimation (\eg, imputation \cite{efron1994missing}).
It aims to ``fix'' recognized incomplete data by replacing missing data with some ``best, reasonable inference'' based on existing data \cite{schafer2002missing}.
This replacement ``breaks" missingness, especially when considering missing data as a type of data \cite{song2018s, schafer2002missing}.
It may bring a false impression of completeness. 
%This could generate serious problems when analyzing such data without \textit{an awareness of missingness}.  
In fact, using such techniques implies that users realize missingness.
%Without an awareness of missingness, how would users know that they should try these methods?
Thus, a successful awareness of missingness is critical for sensemaking activities. 
%%%%%%%%%%%%%%% this was addressed by accepting the changes %%%%%%%%%%%%%%%
%\jian{I reorganized the above few sentences}
 
Visualization can help with missingness awareness for data analytics \cite{andreasson2014effects, eaton2005visualizing}.
With proper visual encodings, missingness gets salient and perceptually attracts user attention.
For example, %as is shown in Figure \ref{fig-empty}, 
given a dataset of connections between two sets of entities, showing it in a matrix with different cell colors (blue indicates the existence of a connection and white means no connection) allows users to see both existing and missing data.
%may be easier for users to realize missing connections than listing all connections in an edge list.
By marrying advanced computation with human cognition with interactive visualizations, visual analytics \cite{keim2008visual} may better handle missingness involved in analyses.
%Visual analytics, ever since its emergent as an interdisciplinary research field \cite{thomas2006visual}, has been getting increasingly important for supporting sensemaking of data.

%\begin{figure}[tb]
%  \centering
%  \includegraphics[width=\columnwidth]{figures/empty-cell}
%  \caption{Compared to listing connections between individual entities (left), it may be easier for users to notice missing connections when showing the same data in a matrix with different cell colors (right).}
%  ~\label{fig-empty}
%  \vspace{-6mm}
% \end{figure}

%%%%%%%%%%%%%%% this was addressed by accepting the comment %%%%%%%%%%%%%%%
%\jian{The following sentence seems useless. We don't need to argue for VA at a vis conference.}
%With interactive visualizations, visual analytics enables users collaboratively working with advanced computational methods to explore solutions to data-driven problems, from simple ones like finding the nearest restaurant for lunch based on the GPS, to complex ones, such as examining medical images or charts for disease diagnose and treatment \cite{shneiderman2013improving}.

%Yet, visual analytics is a powerful approach for supporting sensemaking of data.
Nevertheless, \rv{current understanding of missingness in visual analytics seems scattered and primarily focusing on data values (\eg, missing data \cite{fernstad2019definitionofmissingness}).
However, missingness can be more complex \cite{schafer2002missing}}, when analysts make sense of various data and with different goals. 
%In this work, we study incompleteness from a sensemaking perspective. 
We aim to establish a systematic understanding of missingness in visual analytics and pave the road for future research using visual analytics to handle issues related to or caused by the missingness.

%Questions like, is missingness only related to data; if data is complete, can people still perceive missingness, 
%or does any other possible missingness exist, 
%cannot be clearly answered.
%Due to this, our explorations on using visual analytics to handle issues related to or caused by missingness are limited.
%Moreover, without a systematic understanding of missingness, even for a sensemaking process supported with visual analytics, missingness can still exist and remain unnoticed.

To fulfill this, as an initial step, %for better developing visual analytics techniques to address missingness, 
we consider missingness %that may exist 
in a sensemaking process with visual analytics
%We believe that an in-depth and systematic view of missingness helps inform the design of techniques for handling it. 
%This framework considers missingness 
from two aspects: \textit{data-centric} and \textit{human-centric}.
The former regards the information to be analyzed by users and the latter highlights how users perceive the information.
Specifically, the \textit{data-centric} aspect regards missingness in three data-related categories: \textit{data composition}, \textit{data relation} and \textit{data usage}. 
The \textit{human-centric} aspect considers missingness in three perception-oriented levels: \textit{observed-level}, \textit{inferred-level} and \textit{ignored-level}.
%The two aspects 
They correspond to the two key parties in visual analytics: computation and human cognition, combined together by interactive visualizations.
\rv{Computation can help discover data-related missingness \cite{baraldi2010introduction} (\eg, missing relationship detection \cite{zhao2019missbin, destandau2021missing, zhao2020understanding}).
%Moreover, how data is visualize 
Visualization can impact user awareness of missingness (e.g., missing data) and judgement on data quality \cite{song2018s}.}
Our considerations enable a systematic way of further studying and handling missingness in visual analytics.
\rv{Based on them, visualization design needs to help reveal data-centric missingness and improve user awareness of missingness (e.g., moving from the ignored-level to the observed-level).}
We hope this work can draw attention to future exploration of the design space of visualizing missingness and studying insights from incomplete data in visual sensemaking.
%We hope this work can draw attention to future studies on exploring the design space of visualizing missingness and investigating possibly usable insights from missingness.

\section{A Data-Centric View of Missingness}
\label{data-missingness}

A data-centric view considers missingness in three data-related categories: \textit{data composition}, \textit{data relationship} and \textit{data usage}.
In this section, we first introduce our notion of data composition and data relationship, and then discuss the data-centric view of missingness.

\subsection{Data Composition and Data Relationship}
\label{data-define}

The composition of data can include three major components~\cite{chen1988entity, codd1970relational}: \textit{entity}, \textit{attribute} and \textit{value}.
An entity is a data item, which is a basic unit encoding a piece of information. 
An attribute is a specification that describes an entity, and an entity can have multiple attributes. 
A value reveals how an entity performs on an attribute. 
A \textit{dataset} can be considered as a collection of entities, described by one or multiple attributes with specified values. 
%For example, in a car dataset, each car is a data entity.
%Attributes that describe each car include duration and height, with values such as 1080px and 960px. 
These components have been used for database management, as a relational model \cite{codd1970relational}, which organizes data in tables. 
Each row is an entity, each column corresponds to an attribute, and each cell includes a value. 
%Moreover, data can be considered as objects and each corresponds to a data entity, consisting of attribute and value pairs.

The relationship between data entities can be described using the concept of \textit{entity set}.
An entity set refers to a set of \textit{unique} data entities that share the same attribute(s) (e.g., a collection of photos or a list of locations). 
For two entity sets $X$ and $Y$, a relationship between them, $R (X, Y)$, is a subset of their Cartesian product $X \times Y$. 
When $R (X, Y)$ is not empty, we say $X$ is related to $Y$. 
Otherwise, we say that $X$ is independent from $Y$. 
The relationship $R$ can be determined by using data values of selected data attributes.
Moreover, for different usage scenarios, the relationship $R$ can be determined differently.
For example, in cyber security, $R$ may be defined as communication between computers and web URLs; while in bioinformatics, $R$ may be determined based on expressed genes under conditions.

\subsection{Data Composition Missingness}

\begin{figure}[tb]
  \centering
  \includegraphics[width=\columnwidth]{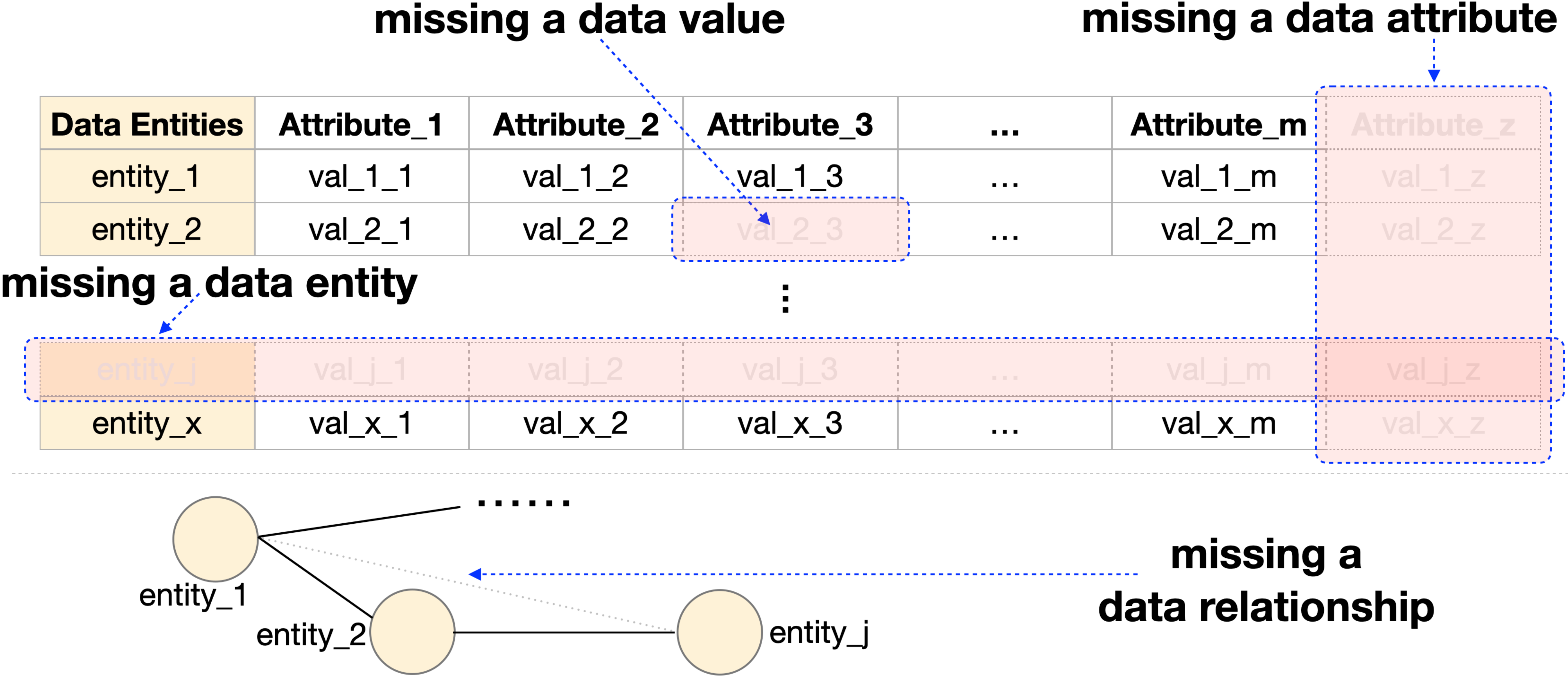}
  \vspace{-7mm}
  \caption{Examples of data composition missingness (top), including: 1) missing data entity, 2) missing data attribute, and 3) missing data value, and data relationship missingness (bottom).}
  ~\label{fig-composition}
  \vspace{-8mm}
\end{figure}

%\begin{figure}[tb]
%  \centering
%  \includegraphics[width=\columnwidth]{figures/relationship}
% \vspace{-7mm}
%  \caption{An example of missing a data relationship.}
%  ~\label{fig-relationship}
%  \vspace{-8mm}
%\end{figure}

Based on the data composition discussed in Section \ref{data-define}, three types of missingness exist: 1) \textit{missing data entity}, 2) \textit{missing data attribute} and 3) \textit{missing data value}.
Figure \ref{fig-composition} (top) gives an example of them.

\textbf{Missing data entities} highlights the absence of data entities. 
It is similar to missing observations \cite{santos2019generating}.
It may result from errors in data collection \cite{allison2001missing}.
For example, in fitness tracking devices, some sensor data might not be recorded due to network connection failures.
%For example, due to the failure of network connections, sensor data cannot be successfully recorded.
%When data entities are lost, their associated attributes and values are also missing. 

\textbf{Missing data attributes} refers to the loss of possibly useful data attributes.
It may come from an ill-defined problem space that leads to problems in data collection mechanisms \cite{pigott2001review} or decisions made by considering some practical concerns.
For example, when designing a logging system of a cloud application, some user behaviors could be overlooked \cite{sun2016designing} or intentionally untracked due to an expensive cost; when creating a survey, researchers may fail to include all related questions or leave some questions unasked for privacy concerns.
For such reasons, even a data collection process runs successfully, some data attributes can still be missing.

\textbf{Missing data values} emphasizes the loss of data values, which is often called missing data for short. %(often named as missing data) 
Compared to the other two, it has drawn the most attention and been heavily studied \cite{song2018s, fernstad2019definitionofmissingness, fernstad2014visualanalyticsofmissingdata}.
Specifically, based on the distribution of data values, missing data values can be categorized into the following three groups \cite{rubin1976}:
% based on their relationship with the data values of observed data entities: 
%1) \textit{missing completely at random (MCAR)}, 2) \textit{missing at random (MAR)} and 3) \textit{missing not at random (MNAR)} \cite{rubin1976}.

\begin{compactitem}%[$\circ$]
\item{\textit{Missing completely at random (MCAR)}: missing data is assumed to not have any underlying mechanism and thus exhibits no relationships with either existing data or other missing data.}
\item{\textit{Missing at random (MAR)}: assumes dependencies on observed values (i.e., existing data), but assumes no underlying relationships between the missing values themselves.}
\item{\textit{Missing not at random (MNAR)}: is the most restrictive, which requires dependencies between missing values.}
\end{compactitem}

\rv{
%Matrix-based visualizations have been used to help reveal them \cite{andreasson2014effects, fernstad2014visualanalyticsofmissingdata, fernstad2019definitionofmissingness}.
Successfully recognizing them can help one understand data under analysis and reasonably generate missing ones for augmentation, if needed (e.g., training machine learning models) \cite{van2001art}.
}
Besides real absent values, missing values has a special case, named as \textit{disguised missing} \cite{pearson2006problem}, in which the value is present but not accurate.
For example, a user leaves a default value (e.g., May 1st for a birthday) for privacy concerns.
In this case, the value is not absent, but instead a default one that may not match the truth.
Thus, in the cases of disguised missing, while values are present, the true information that data collectors need remains missing, as it is intentionally hidden.

\subsection{Data Relationship Missingness}

Data relationship missingness means the absence of relations among data entities.
From a graph perspective, it emphasizes the lack of links among nodes in a graph (Figure \ref{fig-composition} (bottom)).
\rv{This means that for a given set of data entities, some connections between them do not, or with a low probability, exist}.
Missing relationships among data entities may either result from errors in a data collection process or be a reflection of algorithmic results of data relationship discovery.
For example, there is no link between two bank accounts due to the loss of an intelligence report; \rv{or the connection between two persons is computed as a very low probability based on word cooccurrence}. 

Missing data relationships can be formalized as the problem of missing links in graphs.
Similar to using the imputation techniques for missing value inferences, based on existing links, missing links can be computationally identified \cite{zhao2019missbin, zhao2020understanding}.
A common goal of such techniques is to find potentially useful missing links (e.g., serving as a bridge that connects two communities in a social network), and further fix and verify them by adding back the lost ones \cite{zhao2020understanding}.

\subsection{Data Usage Missingness}

\begin{figure}[tb]
  \centering
  \includegraphics[width=\columnwidth]{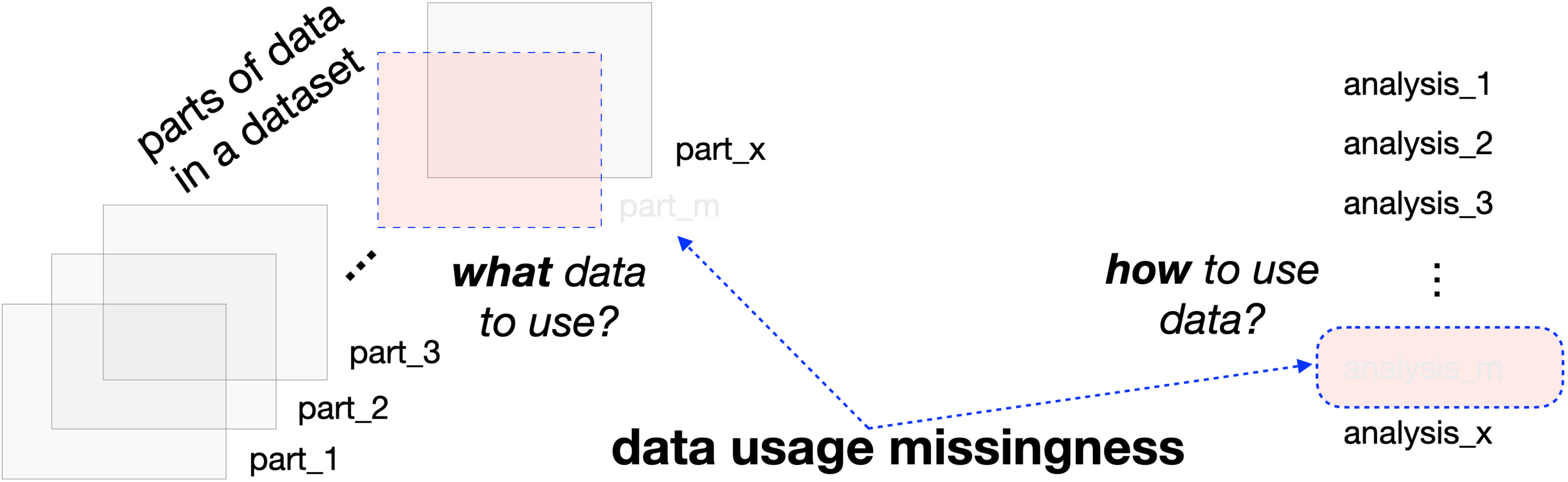}
  \vspace{-6mm}
  \caption{Two possible missingness in data usage: 1) missing data selection, and 2) missing analytical method selection.}
  ~\label{use-incomplete}
  \vspace{-8mm}
\end{figure}

The utility of data for sensemaking activities involve two key types: 1) \textit{data selection} and 2) \textit{analytical method selection}.
The former refers to which parts of a given dataset will be selected for analysis.
The latter means which analytical methods will be picked and applied to the selected data.
Missingness in data usage can happen in both activities due to uncertainties and selection biases \cite{heckman1990varieties, wall2019toward}).
%Due to various reasons (e.g., uncertainty and selection bias \cite{heckman1990varieties, wall2019toward}), missingness in data usage may appear at both of them.

\textbf{Missing data selection} reveals that not the whole dataset is selected for analysis.
For example, to find similar cars, 4 out of 100 attributes are selected and the rest remains unused; or instead of using all data entities, a dataset is sampled and then analyzed.
When selected attributes or samples of data entities are not representative (e.g., stratified sampling \cite{trost1986statistically}), missing data selections occurs.

\textbf{Missing analytical method selection} reveals that not a full set of possibly applicable analytical methods is selected and tested.
For real-world problems, it is not easy or sometimes even impossible to identify a complete set of analytical methods.
Thus, this missingness highlights that performed analyses are not sufficient, which calls for further or alternative analysis (i.e., multiverse analysis \cite{dragicevic2019increasing}).
For example, to explore similar cars, only one centroid-based clustering method (i.e., k-means clustering) is used, but other applicable clustering techniques that may work are not studied.

%A dataset is a collection of entities meant to sample entity sets that occur in a population. Data utilization incompleteness occurs when the relationships and or values of the entities in a sample, or data set, are significantly different than the values of the entities in the entity set. If the sample set is not a subset of the population set or the sample set is a biased set of values, incomplete data utilization occurs. 

%Non-representative sampling is also an example of incomplete data utilization. A sample can be non-representative of the population it is meant to describe, either by an error attributable to selection or collection, or purposefully. In some cases, non representative samples are a result of errors in the collection process \cite{rubin1976}, but in some cases the analyst specifically chooses a non representatives sample \cite{goel2015FastCheap, trost1986statistically}. Both are examples of incomplete data utilization.

\section{A Human-Centric View of Missingness}
A human-centric view treats missingness at three levels: 1) \textit{observed}-level, 2) \textit{inferred}-level and 3) \textit{ignored}-level.
They reveal how data-centric missingness discussed in Section \ref{data-missingness} is perceived by people.

\subsection{Observed Missingness}
Observed missingness refers to that users can directly perceive missingness.
It indicates that the visibility of missingness is high, and users can easily notice it.
For example, a user quickly realizes that a data value is missing, after she sees an empty cell in a table; or by checking and following ribbons in a parallel sets \cite{kosara2006parallel}, a user finds that there is no connection between two categorical data entities \cite{convertino2019method}.

\begin{figure}[tb]
  \centering
  \includegraphics[width=\columnwidth]{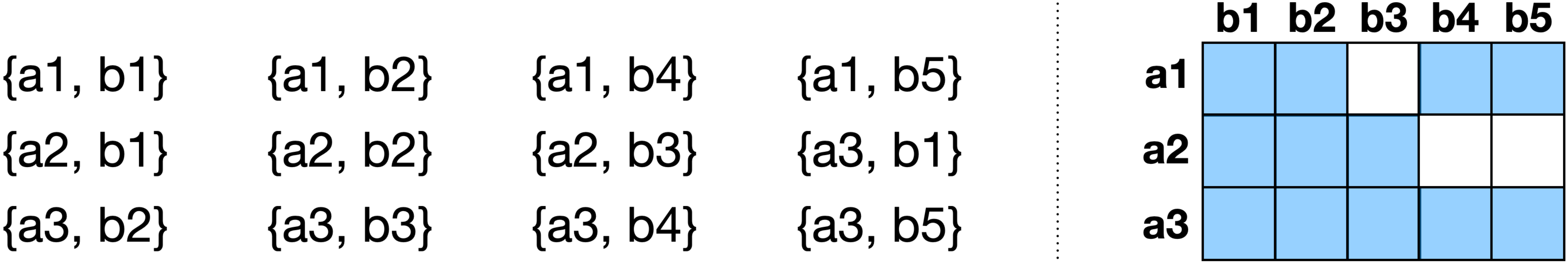}
  \vspace{-7mm}
  \caption{Compared to listing connections between individual entities (left), it is easier for users to see missing connections when showing the same data in a matrix with different cell colors (right).}
  ~\label{fig-empty}
  \vspace{-8mm}
 \end{figure}

As the visibility of missingness is affected by the way that data is represented, observed missingness relies on visual context, in which data is encoded by certain visualizations.
Different visual encodings can impact how easily users can observe missingness.
For example, as is shown in Figure \ref{fig-empty}, it is easier for users to see missing links by looking at a matrix than checking the same data displayed as lists of node-pairs.
Thus, for observed missingness, users can verify their perceived missingness by referring to certain given visual context (e.g., pointing to an empty cell).

\subsection{Inferred Missingness}
Inferred missingness refers to that the visibility of missingness goes low or missingness even gets invisible, so it is impossible for users to directly observe missingness.
However, via an investigation with given data, users can infer the possible existence of missingness.
For example, by reading the following four intelligence reports that are modified based on the Sign of the Crescent dataset \cite{hughes2003discovery}:

{\small\textit{``Report on 04/24/2003. Phone calls on 22 April, 2003 made from 703-659-2317 to the numbers: 804-759-6302 and 804-774-8920. A translation of this message reads: `I will be in my office on April 30 at 9:00AM. Try to be on time'.}}

{\small\textit{Report on 01/11/2003. Abdul Ramazi is the owner of the Select Gourmet Foods shop in Springfield Mall, Springfield, VA., with a phone number 703-659-2317.}}

{\small\textit{Report on 03/18/2003. A check with mobile phone providers shows that a Sprint cell phone 804-774-8920 is registered in the name Mukhtar Galab.}}

{\small\textit{Report on 04/14/2003. The contact given by Faysal Goba was: 1631 Capitol Ave., Richmond VA; phone number: 804-759-6302. From an interrogation of a cooperative detainee in Guantanamo. Detainee says he trained daily with a man named Faysal Goba at an Al Qaeda explosives training facility in the Sudan in 1994."}}

One may infer that the three persons, \texttt{Abdul Ramazi}, \texttt{Mukhtar Galab} and \texttt{Faysal Goba} may collude suspicious activities together. 
However, it seems missing in the given reports as it was not explicitly reported.
Compared to observed missingness, inferred missingness may not be easily verified.
Thus, observed missingness seems more confirmative, while inferred missingness is more hypothetical, which suggests and is closely related with implicit uncertainty \cite{mccurdy2018framework, panagiotidou2021implicit}.

\subsection{Ignored Missingness}
Ignored missingness indicates no observation nor awareness of missingness and the presence of possible missingness is not considered.
It may occur for two reasons.
First, the visibility of missingness is too low to raise user awareness.
For example, in Figure \ref{fig-empty}, a user may never realize that missing edges exist after looking at lists of node-pairs.
Second, due to some biases or the impact of cognitive capture (or tunneling) \cite{simons1999gorillas}, users turn a blind eye to possible missingness.
For example, to explore possible treatment for a disease, all effort has been put on the group of people who have been infected by the disease, while the uninfected group never gets any attention.

While it cannot be completely avoided in the analysis, ignored incompleteness, if identified, can bring critical values for sensemaking activities \cite{pirolli2005sensemaking}.
%While ignoring missingness seems a phenomenon that occurs in user analyses, the ignored missingness, if successfully recognized, can bring critical values for sensemaking activities \cite{pirolli2005sensemaking}.
Based on this, we consider that ignored missingness is similar to the concept of \textit{white space} (also named as \textit{opportunity space}) discussed in the business domain \cite{johnson2010seizing}.
The white space suggests new leads for possible growths of a business.
For example, the customers of a credit card product fall into two major age groups 25-35 and 50-70. 
The gap between 35 and 50 is a white space. 
It indicates that the current product seems not attractive for the age group 35-50, given the lack of users. 
Thus, this white space implies an opportunity to design a different credit card product with new awarding features for competing in the market of the missing age group.
While a white space can bring useful values to business, it is usually difficult to capture and may easily slip one's attention  \cite{johnson2010seizing} (e.g., the missing age group catches no attention at all).  

In summary, observed missingness takes the least amount of user effort to perceive possible missingness; while for ignored missingness, users are not aware of the existence of missingness in the whole sensemaking process.
Moreover, for inferred missingness, users can realize missingness but it takes more effort.

\section{Handling Missingness: Visualization Roles}
Based on the data-centric and human-centric perspectives of missingness mentioned before, here we discuss four possible roles of visualizations for handling missingness.
The first and second roles highlight supporting the detection of data-centric missingness.
The other two roles aim to improve user awareness of data-related missingness.
In summary, visualizations can help to uncover the data-centric missingness and improve their expressiveness, so they become more visible and accessible to users.

\subsection{Bridging Existing Data and Missing Data}
Visualizations play a key role of bridging the gap between existing data and missing data.
If we consider existing data as a visible land and missing data as an invisible world, a usable bridge connecting them is critical to enable users to explore and walk into the invisible part from the visible one.
This is because users need existing data as a landing point before digging into the data-centric missingness.
However, to enable the analytical transition from the existing data to the missing part, users need the support of necessary information hints or leads, which can be provided by visualizations \cite{chi2001using}.

To establish such a bridge, a commonly used strategy is \textit{space-filling} that reveals missingness as empty (e.g., an empty space in a bar chart \cite{song2018s}), gap (e.g., broken lines in a line chart \cite{song2018s}), or different-looking space (e.g., a matrix with different colored cells \cite{zhao2019missbin, fernstad2019definitionofmissingness}).
The focus of such visualization techniques are on the existing data.
As the present data is mapped to certain visual encodings, possible missingness gets visible.
Looking at a visually salient space, users can be aware of data-centric missingness.
Thus, visualized existing data serves critical and usable visual context that enables users to identify data-related missingness.

\subsection{Supporting the Analysis of Analytic Provenance}
Visualizations can serve a usable solution to understand and audit analytical provenance \cite{ragan2015characterizing, li2020crowdtrace}, which is helpful to address incompleteness in data usage.
It is challenging for users to keep tracking the process of their analyses.
In a sensemaking process, some parts of data may not receive enough attention and users may miss one or several possibly applicable methods unintentionally.
To help avoid such data usage missingness, visualizations can be used to support tracking analytical provenance and further analyze it.

To help identify missingness in data usage, two key aspects need to be considered: 1) the selected, investigated, derived and newly generated data, and 2) the method or process applied to such data.
They, respectively, correspond to the provenance of data and process \cite{simmhan2006performance}.
Visualizing them offers a way of analyzing analytic provenance.
By checking such visualizations, users may notice missingness in data usage and further overcome limitations of their analyses.

\subsection{Improving Awareness: from Ignoring to Observing}
From a perceptual-oriented perspective, a key role of visualization is to prevent users from falling in the trap of ignoring data-centric missingness.
The presence of missingness can become more visible to users via the usage of visualizations than without them, so it is more likely for users to be aware of missingness.
This implies that using visualizations can improve the \textit{expressiveness} of data-centric missingness.
The higher such expressiveness goes, the easier it is for users to observe possible missingness.
Thus, using visualizations to handle data-centric missingness attempts to move forward from ignoring missingness to being able to observe it.

Proper visual encodings can direct user attention to data-centric missingness (e.g., missing data values) \cite{song2018s}, which may otherwise be ignored by users.
The data-centric missingness is often unknown to users at the initial analysis stage, unless they are informed.
Thus, a sensemaking process with data-centric missingness is exploratory in nature and the original analysis goal may not consider missingness at all.
However, by referring to visualizations used in a sensemaking process, users may realize the existence of missingness, which could happen at an ``aha" moment \cite{mai2004aha}.
This matches both the \textit{spontaneous} insight \cite{chang2009defining} of visual analytics and one of the key characteristics of visualization insight -- \textit{unexpected} \cite{north2006toward}. 

\subsection{Scaffolding Missingness Inference}
Visualizations offer a usable mean to scaffold missingness inference.
Different from the other two perceptual levels of missingness (i.e., observing and ignoring missingness), inferring missingness requires more user effort, as possible missingness is not directly revealed but somehow can be inferred with enough cognitive effort.
This can be supported by using visualizations.
In this case, instead of merely encoding missing parts of data or existing parts for the purpose of indicating the ``hole" in data, visualizations may focus on displaying either the connections across different parts of data or the provenance of a sensemaking process.
These help users to infer possible existence of data-centric missingness.

As inference is a reasoning process, instead of a static stage, using multiple types of visualizations and fusing information across them may help with missingness inference.
For example, by examining links in a social network graph, checking related organizations, and reading relevant reports, users may infer that two suspects colluded some threats together, which was never reported in a given dataset.
\rv{This visualization-aided inference making may help one identify the data-centric missingness, such as checking multiple visualizations in a dashboard about the usage of a cloud application to find attributes of user interactions that have not been logged before \cite{sun2016designing}.}
%Thus, visualizations can play an important role in scaffolding missingness inference.%, but it has more complex needs for the design of visualizations.

\section{Discussion and Conclusion}
While handling missingness remains a challenging problem in sensemaking, as an initial step, we present considerations that may help to systematically study missingness in visual analytics.
They highlight considering missingness from two key perspectives: \textit{data-centric} and \textit{human-centric}.
The former regards missingness in three data-related categories: \textit{data composition}, \textit{data relationship} and \textit{data usage}. The latter focuses on the human-perceived missingness at three levels: \textit{observed}-level, \textit{inferred}-level and \textit{ignored}-level.
Based on the considerations, we discuss four possible roles of visualizations for helping to handle missingness in a sensemaking process.
While they help to lay a preliminary theoretical foundation that aims to systematically consider missingness in visual analytics, to handle missingness in practice, there are four research themes that are worthy of future studies: 1) missingness detection, 2) missingness visualization, 3) missingness insight, and 4) possible relations between missingness and uncertainty.

\subsection{Detecting Missingness}
%Missingness detection is a key research direction. 
%Regarding the three categories of data-centric missingness, missingness in data composition seems easy to detect. 
Missingness detection lays the foundation for effective data analysis.
Detecting missingness is not as simple as it looks like. 
Unless there is a clear detection goal or some evidence that reveals something is missing (e.g., data value), detecting missingness is fundamentally attempting to address an \textit{unknown unknown} problem \cite{matta2018making}.
This brings a deeper question: %besides focusing on the exploration of detection techniques.
how can we help users know which types of data-centric missingness (e.g., missing data attributes, missing data relationships, or missing data selections in the usage) exist?
This is essential as it sets the detection goal.
If users were not clear about this, it would be hard for them to further explore and work on detection methods.
Also, a sensemaking process can have multiple types of data-centric missingness. 
For example, a vulnerable system with an interrupted network connection and a problematic logging mechanism can lead to missing both data values and attributes.
For such cases, detecting missingness is even challenging.
Our considerations presented in this work may help to clarify detection goals.

\subsection{Visualizing Missingness}
The design of missingness-oriented visualizations remains an under-explored direction.
Prior work has investigated visual encodings for missing values \cite{song2018s, fernstad2019definitionofmissingness, fernstad2014visualanalyticsofmissingdata} and missing links \cite{zhao2019missbin}. 
However, the design space of visualizing missingness can be broader, especially considering that there are different categories of data-centric missingness and they may need different visual encodings.
As studied in \cite{song2018s}, even for the same type of missingness, different visual encodings can be designed, which further impacts user-perceived data quality.
How to formalize the design space of missingness visualizations still needs further explorations.
Furthermore, considering the evaluation of missingness visualization, how and if possible can we measure the expressiveness of visual encodings for data-centric missingness?
It enables comparing different designs for visualizing missingness, which can be helpful to support making design decisions.
The perceptual-perspective discussed in our framework may help to derive usable measures.

\subsection{Discovering Insights from Missingness}
Studying possible insights that users gain from missingness in sensemaking is a highly sought-after research challenge.
Missingness can be considered as a type of ``data" \cite{song2018s} from which users can gain usable insights (e.g., using partial bipartite graphs for performing grouping tasks \cite{sun2018effect}). 
This turns missingness from being considered as dirty \cite{kim2003taxonomy} to usable.
%Since missingness can be considered a type of ``data" \cite{song2018s}, studying useful insights from it potentially turns missingness from commonly considered dirty \cite{kim2003taxonomy} to usable.
For example, in an intelligence analysis, a missing link between two suspects may drive the subsequent analysis towards an investigation of any possible connections between them \cite{sun2014role}.
While this is a simple example, it shows that missingness can be used in a sensemaking process.
The insights derived from missingness may depend on an application domain and different types of data-centric missingness may bring different insights.
Moreover, insights discovered from missingness, if possible, via using visual analytics, may enlarge the set of characteristics of visualization insight \cite{north2006toward}.
This may further broaden our understanding of evaluating visualizations by considering the value of missingness.

\subsection{Relating Missingness with Uncertainty}
An in-depth understanding of possible relations between missingness and uncertainty remains under-explored.
For one thing, data-centric missingness brings uncertainty about data and analysis.
In practice, uncertainty resulted from data loss, regardless of intentionally or not, may not be well-resolved, as it may be impossible to collect the truth.
Thus, would user awareness of data-centric missingness help with uncertainty-based decision making remains an unanswered question.
For another, uncertainty may impact user awareness of missingness.
While uncertainty may occur due to a variety of reasons \cite{hullman2019authors}, would some level of visually expressed uncertainty impact user awareness or perception of missingness?
Specifically, after being exposed to some visualized uncertainty, would users associate this with missingness or would this help direct users to starting thinking or inferring data-centric missingness?
Answers to such questions can help enrich the design space of uncertainty visualizations and advance knowledge about sensemaking under uncertainty and missingness.
To find such answers, further studies are needed.

In summary, we present considerations of missingness in visual analytics from two aspects: \textit{data-centric} and \textit{human-centric}, which offers a possible way of further systemically studying missingness.
We hope this work can draw attention to future studies on visual sensemaking with missingness.


\acknowledgments{
This research is supported in part by NSF Grants IIS-2002082 and DMS-2152070, the Research and Artistry Opportunity Grant from Northern Illinois University, and the University of Waterloo International Research Partnership Grants (IRPG).}

\bibliographystyle{abbrv-doi}

\bibliography{paper}
\end{document}